# Ultrathin graphene-based membrane with precise molecular sieving and ultrafast solvent permeation

Q. Yang[1,2,†], Y. Su[1,2,†,*], C. Chi[1,2,†], C. T. Cherian[1,2], K. Huang[1,2], V. G. Kravets[3], F. C. Wang[4], J. C. Zhang[5], A. Pratt[5], A. N. Grigorenko[3], F. Guinea[3,6], A. K Geim[3], R. R. Nair[1,2*]

[1]National Graphene Institute, University of Manchester, Manchester, M13 9PL, UK.

[2]School of Chemical Engineering and Analytical Science, University of Manchester, Manchester, M13 9PL, UK.

[3]School of Physics and Astronomy, University of Manchester, Manchester M13 9PL, UK.

[4]Chinese Academy of Sciences Key Laboratory of Mechanical Behavior and Design of Materials, Department of Modern Mechanics, University of Science and Technology of China, Hefei, Anhui 230027, China.

[5]Department of Physics, University of York, York YO10 5DD, UK.

[6]Imdea Nanociencia, Faraday 9, 28015 Madrid, Spain.

†These authors contributed equally to this work.

*yang.su@manchester.ac.uk & rahul@manchester.ac.uk

**Graphene oxide (GO) membranes continue to attract intense interest due to their unique molecular sieving properties combined with fast permeation rates[1-9]. However, the membranes' use has been limited mostly to aqueous solutions because GO membranes appear to be impermeable to organic solvents[1], a phenomenon not fully understood yet. Here, we report efficient and fast filtration of organic solutions through GO laminates containing smooth two-dimensional (2D) capillaries made from flakes with large sizes of ~ 10-20 μm. Without sacrificing their sieving characteristics, such membranes can be made exceptionally thin, down to ~10 nm, which translates into fast permeation of not only water but also organic solvents. We attribute the organic solvent permeation and sieving properties of ultrathin GO laminates to the presence of randomly distributed pinholes that are interconnected by short graphene channels with a width of 1 nm. With increasing the membrane thickness, the organic solvent permeation rates decay exponentially but water continues to permeate fast, in agreement with previous reports[1-4]. The application potential of our ultrathin laminates for organic-solvent nanofiltration is demonstrated by showing >99.9% rejection of various organic dyes with small molecular weights dissolved in methanol. Our work significantly expands possibilities for the use of GO membranes in purification, filtration and related technologies.**

Membrane-based technologies enable efficient and energy-saving separation processes which could play an important role in human life by purifying water or harvesting green energy[10,11]. Recently, it was shown that molecular separation processes could benefit from development of graphene-based membranes[2-4] that show tunability in pore size[8,12-15] and ultimate permeance[15] defined by their thinness. In particular, GO-based membranes are considered to be extremely promising for molecular separation and filtration applications due to their mechanical robustness and realistic prospects for industrial scale production[2-4,7,9]. A considerable progress in nanofiltration through GO membranes[2-4,16] was achieved mainly for water (due to its ultrafast permeation[1-4]) while organic-solvent permeation has received limited attention. This disparity is rather surprising as organic solvent nanofiltration (OSN) attracts a tremendous interest due to its prospective applications in chemical and pharmaceutical industries[11,17-20]. The development of novel inorganic membranes for OSN is particularly vital because of the known instability of many polymer-based membranes in organic solvents. The possible lack of motivation for exploiting graphene-based membranes for OSN could have come from the previous reports on impermeability of organic solvents through sub-micron thick GO membranes that remained highly permeable for water[1,2,21]. Although some latest studies report the swelling of GO membranes in organic solvents and, accordingly, indicate permeability of organic molecules even through thick GO membranes[22,23], this seems inconsistent with the previous reports[1,2,21] and could be explained by the presence of extra defects that produce a molecular pathway. In an another work[24] OSN was performed using a solvated reduced GO-polymer composite membrane and only achieved a molecular sieve size of ≈ 3.5 nm due to the larger nanochannels in the membrane than that of pristine GO membranes[1,2,5]. Molecular rejection for the above membranes involves charge specific separation rather than the physical size cutoff. Membranes with Angstrom size precise sieving along with high organic solvent permeance are of great interests for OSN technology, however, such demostration is still lacking. In this report, we investigate permeability and sieving properties of ultrathin GO membranes with respect to organic solutions using an improved laminar structure and demonstrate the membranes' potential for OSN.

The preparation of GO membranes used in our work is described in Methods. Figure 1 shows the scanning electron microscope (SEM), atomic force microscope (AFM) images and X-ray diffraction (XRD) of the studied GO membranes. Short duration ultrasonic exfoliation and a stepwise separation (Methods) were used to obtain large GO flakes (lateral size *D* of 10 – 20 μm) with a relatively narrow size distribution (supplementary Fig. 1). The membranes prepared from these large GO flakes are referred to as highly laminated GO (HLGO) membranes due to their superior laminar structure. They show a narrow XRD peak (full width at half maximum of 0.4 degree) as compared to 1.6 degree for the standard GO membranes prepared from smaller flakes (*D* ~ 0.1 – 0.6 μm). Below the latter are referred to as the conventional GO (CGO) membrane. The narrow X-ray peak for HLGO laminates suggests the importance of the GO flake size for the alignment process, which can be attributed to stronger interlayer interaction between larger overlapping areas[25]. The stronger interactions could further assist to eliminate the occasional wrinkles and corrugation found in the CGO membranes[2,3], and this could lead to achieving smoother 2D capillaries in HLGO membranes.

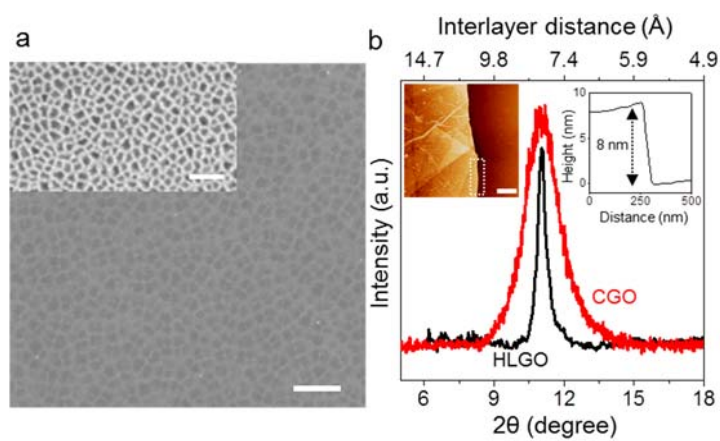

**Figure 1| Ultrathin HLGO membrane.** (a) SEM image of an 8 nm thick HLGO membrane on an Anodisc alumina support. Scale bar, 1µm. Inset: SEM image of bare alumina support. Scale bar, 500 nm. (b) X-ray diffraction for HLGO and CGO membranes. Inset (left): AFM image of HLGO membrane transferred from an alumina substrate to a silicon wafer. Scale bar, 500 nm. Inset (right): The height profiles along the dotted rectangle.

To probe molecular sieving properties of HLGO membranes, we first performed vacuum filtration of aqueous solutions of several salts and large molecules through HLGO membranes (Methods). Figure 2a shows the molecular sieving properties of an 8 nm thin HLGO membrane. Similar to micron-thick GO membranes[5], HLGO membranes also block all ions with hydrated radii larger than 4.5 Å. We emphasize that no molecular sieving was observed in similar experiments but using CGO membranes with thickness of 8-50 nm (Fig. 2a inset). Hence, the ultra-sharp sieving cut-off can be achieved in HLGO membranes that are more than two orders of magnitude thinner than conventional membranes showing same sieving properties[5]. This drastic improvement can be attributed to the highly laminated nature of our HLGO membranes. We failed to observe a cut-off in sieving only for the membranes thinner than 8 nm, which sets a minimum thickness for HLGO membranes used in this study.

Ultrahigh permeance to fluids may occur in ultrathin membranes due to a decreased molecular permeation length[6,15]. To further evaluate liquid permeance of HLGO membranes, we have performed vacuum filtration and dead-end pressure filtration (supplementary section2) experiments with water and a wide range of organic solvents using only 8 nm thick membranes. All the permeance values were recorded after reaching a steady state condition, typically achieved within 30 minutes. The liquid flux is found to be linearly proportional the differential pressure (ΔP) across an HLGO membrane (Fig. 2b inset). The permeance for various solvents as a function of their inverse viscosity ($1/\eta$) is shown in Fig. 2b. In contrast to much-thicker GO membranes that exhibit ultrafast water permeation and impermeability for organic solvents[1], our HLGO membranes are highly permeable to all tested solvents. The highest permeance is observed for solvents with the lowest viscosity. For example, hexane shows permeance of ~18 $Lm^{-2}h^{-1}bar^{-1}$, i.e, a permeability of ~144 $nm \cdot Lm^{-2}h^{-1}bar^{-1}$, despite the fact that its kinetic diameter is almost twice larger than that of water[26]. On the contrary, 1-butanol with a kinetic diameter similar to that of hexane[26] but much higher viscosity exhibits the lowest permeance of 2.5 $Lm^{-2}h^{-1}bar^{-1}$. The linear dependence of permeance on $1/\eta$ (see Fig.2b) clearly indicates that the solvent viscosity dictates its permeability and proves the viscous nature of the solvents' flow through HLGO membranes.

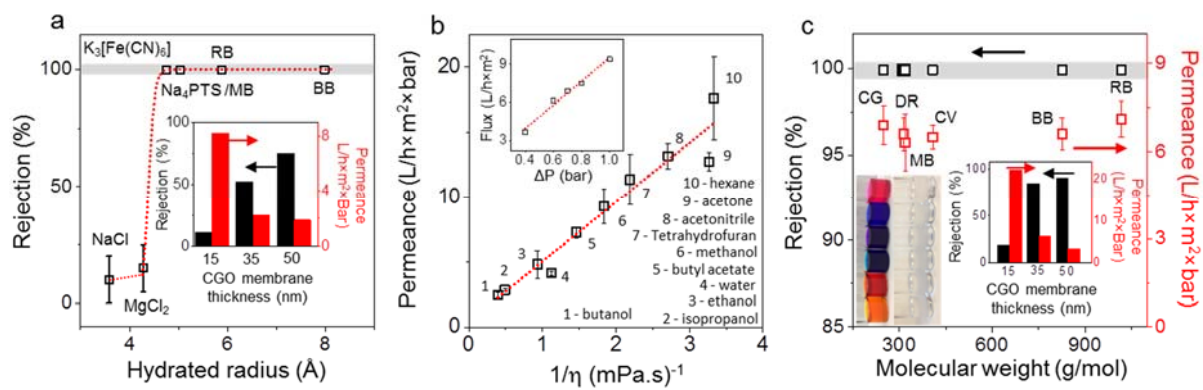

**Figure 2| Molecular sieving and organic solvent nanofiltration through HLGO membranes.** (a) Experiments for salt rejection as a function of ion's hydrated radius (largest ions within the aqueous solutions are plotted). The HLGO membranes are 8 nm thick. The hydrated radii are taken from ref. [5 and 7]. MB- Methylene Blue, RB – Rose Bengal, BB – Brilliant Blue. Inset: MB rejection and water permeance exhibited by the standard GO membrane with different thicknesses (colour coded axes). (b) Permeance of pure organic solvents through an 8 nm HLGO membrane as a function of their inverse viscosity. The used solvents are numbered and named on the right. Inset (top): Methanol permeance as a function of pressure gradient (ΔP). Dotted lines: Best linear fits. (c) Rejection and permeance of several dyes in methanol versus their molecular weight (colour coded axes). The dyes used: Chrysoidine G (CG), Disperse Red (DR), MB, Crystal Violet (CV), BB and RB. Left inset: Photographs of dyes dissolved in methanol before and after filtration through 8 nm HLGO membranes. Right inset: MB rejection and methanol permeance of CGO membrane with different thicknesses (colour coded axes). Note that even though the dye rejection increases and approaches ~ 90% with increasing the CGO membrane thickness their permeance is

significantly lower than 8 nm HLGO membranes. All the error bars are standard deviations using at least three different measurements using different samples. Points within the grey bar in Figs.1a and c show the rejection estimated from the detection limit (supplementary Fig. 4 and Methods).

High permeance of organic solvents combined with accurate molecular sieving makes ultrathin HLGO membranes attractive for OSN[17,18]. To evaluate this potential for applications, we have performed filtration experiments with methanol solutions of several dye molecules. The dye molecule rejection rates for an 8 nm thick HLGO membrane are presented in Fig. 2c. While the permeance was reduced by ~10-30% compared to the pure solvent (which is not unusual for nanofiltration[16,27]), no dye molecule could be detected down to 0.1% (our detection limit) of the feed concentration at the permeate side (Fig. 2b). The observed ~100% dye rejection and fast solvent permeation makes our ultrathin HLGO membranes superior to the state-of-the-art polymeric membranes[17,19]. For example, the highest methanol permeance reported on polymeric membranes[17,19] is ~ 1.6 $Lm^{-2}h^{-1}bar^{-1}$ for 90% RB rejection which is ≈ 5 times lower than the methanol permeance obtained with our HLGO membranes providing ≈ 100% RB rejection. Further comparison of OSN performance of HLGO membrane and the other reported OSN membranes are listed in supplementary section 4. Unambiguously, a high organic solvent permeance along with precise molecular sieving (> 99.9% rejection to the dye molecules) indicates that HLGO membranes could be an outstanding candidate for OSN technology. With the view of practical applications, we have also performed OSN experiments with HLGO deposited on porous polymer (nylon) support (supplementary section 5). The nylon supported HLGO membranes showed nearly the same performance as those on the alumina support. For example, an 8 nm HLGO membrane on nylon showed a >99.9% rejection to MB with ≈ 7 $Lm^{-2}h^{-1}bar^{-1}$ methanol permeance (supplementary Fig. 6). We have also studied the influences of aging and solvent stability of HLGO membrane, which are key parameters for practical applications, on its membrane performance, and found that HLGO membranes are stable in air for more than a year and also highly stable in different solvents (see supplementary section 6).

To elucidate the mechanism of organic-solvent permeation and sieving properties of ultrathin HLGO membranes, we have conducted two sets of additional experiments. First, we have performed XRD for HLGO membranes immersed in different organic solvents, see Fig. 3a. The data clearly indicate that several of the organic solvents, especially polar ones, intercalate between graphene oxide layers and increase the interlayer distance, $d$. However, non-polar solvents, such as hexane, did not produce any increase in $d$. At the same time, hexane was the fastest permeating molecule among the solvents used in this study (Fig.2b). This suggests that permeation through ultrathin HLGO membranes is not dominated by molecular transport through interlayer capillaries[1]. Second, we performed water and organic solvent permeation experiments using HLGO membranes of different thicknesses, $h$. Fig. 3b shows the exponential decay for methanol and hexane permeance as a function of $h$. HLGO membranes with $h > 70$ nm show no detectable solvent permeation, consistent with the impermeability reported for sub-micron thick GO membranes[1]. Using helium and organic vapours, we also observed a similar, exponential decay with increasing $h$ of our HLGO membranes (supplementary section 7). In contrast, water permeance initially decayed exponentially, too, but for $h > 70$ nm it followed a much weaker, linear dependence on $1/h$ (Fig.3b inset).

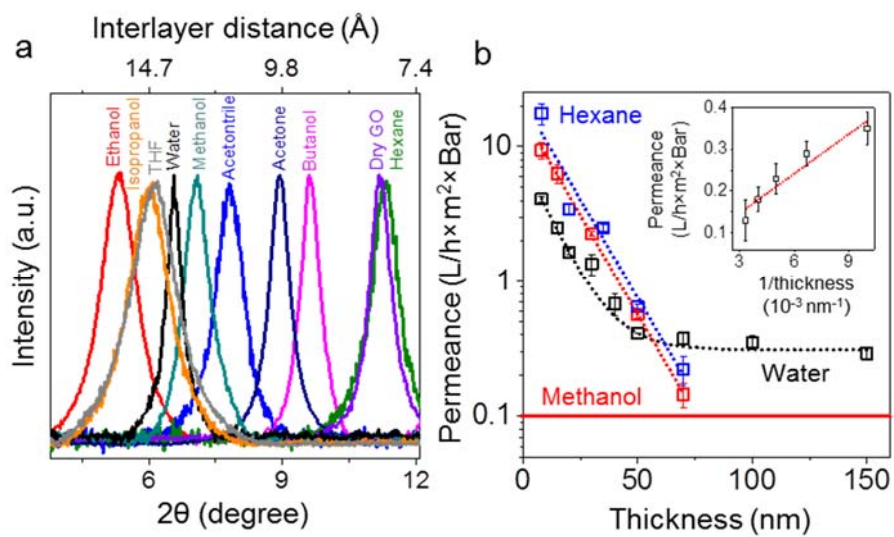

**Figure 3| Probing molecular permeation through HLGO membranes.** (a) X-ray diffraction for 70 nm thick HLGO membranes immersed in various organic solvents (colour coded). (b) Thickness dependence of permeance for methanol, hexane, and water through HLGO membranes (colour coded). Red and blue dotted lines are the best exponential fits. The black dotted curve is a guide to the eye. Inset: Water permeance as a function of inverse thickness for HLGO membranes with thicknesses ≥100 nm. Dotted line: best linear fit. The slope of linear fit provides the water permeability as ≈ 32 nm·Lm$^{-2}$h$^{-1}$bar$^{-1}$. The solid line in the main figure shows the detection limit for methanol and hexane in our experiment. All the error bars are standard deviations using at least three different measurements using different samples.

The exponential decrease of organic-solvent permeance with $h$ is surprising and seemingly contradicts to the viscous flow inferred from the observed $1/\eta$ dependence. Indeed, the viscous flow suggests that the permeance should be proportional to the pressure gradient $\frac{\Delta P}{L}$, where $\Delta P$ is the driving pressure gradient and $L$ is the permeation length (proportional $h$)[28,29]. For example, the linear dependence of water permeance on $1/h$ for the thicker membrane is consistent with the viscous flow. To explain these two functional dependences, we propose two different molecular pathways for permeation through HLGO membranes. The first involves permeation through pin holes (pathway 1) and the second one is through the previously suggested model of a network of graphene capillaries[1,5] (pathway 2).

Pin holes in GO membranes originate from random stacking of individual GO flakes and can also involve nanometre size holes[2] within flakes. At a few nm thicknesses, GO laminates contain many pinholes (supplementary section 8) that pierce through the entire film. Such thin GO films allow relatively easy permeation through pinholes without any atomic-size cutoff observed for thicker laminates. At a certain critical thickness $h_c$, GO films become continuous with all pinholes blocked, as the found onset of atomic-scale sieving indicates. The experiment shows that for HLGO membranes, $h_c$ is ~8 nm. After this threshold, molecular transport is expected to occur in two steps. Liquids continue to rapidly fill the same pinholes but this is not a limiting process. Molecular transport through the entire film becomes limited by the necessity to reach from one pinhole to another, which involves in-plane diffusion between GO sheets. This bottleneck has to involve interlayer diffusion by a distance of the order of the size of GO sheets, which will provide an atomic-scale sieve size for filtration. Assuming that a probability for a molecule to find a pathway through the thinnest continuous membrane with critical thickness $h_c$ is $p$, we can write the probability of transport through a thicker sheet of thickness $h$ as $P = p^N$ where $N = h/h_c$. This can be re-written as $P = \exp[\ln(p)h/h_c]$ and yields the flux $Q \propto \exp(-h/a)$ with $a = h_c/\ln(1/p)$. By definition $p$ should be of the order of ½ because we define it at the threshold, which means that $p \Rightarrow 1$ for $h < h_c$ and $p \Rightarrow 0$ for thicker layers. Therefore, $a = h_c/\ln(1/p) \sim h_c$, in agreement with exponential fit in Fig. 3b. This proposed model could also explain why the molecular sieve size of $\approx 4.5$ Å is not preserved in thin CGO membranes where the smaller flake size increased the critical thickness and weakened the interlayer alignment.

The deviation of water permeance from the exponential decay and it's faster transport at large $h$ can be understood by considering the molecular pathway 2 where the permeation occurs through the graphene capillaries[1,5] that develop between GO sheets (typically, an area of 40-60% remains free from functionalization[30,31]). The permeation through the pathway 2 is primarily restricted by the hydraulic resistance due to a large $L$ ($\frac{D}{d} \times h$)[1]. However, water permeation through these capillaries experiences three orders of magnitude enhanced flow due to the large slip length[1,8,29] and therefore effectively reduces the flow resistance. This suggests that with increasing $h$, the exponentially growing flow resistance for water in the pathway 1 could be overcome by the lower flow resistance in the pathway 2 due to the large slip length, consistent with the deviation from the exponential decay observed above ~ 50 nm in Fig. 3b. The linear dependence of water permeance on $1/h$ for the HLGO membranes with a thickness larger than 70 nm (Fig. 3b inset) further proves that the flow through the thicker membranes predominantly occurs through the graphene capillaries[29]. In contrast, for organic solvents, the experimentally undetectable permeation for $h > 70$ nm indicate that the permeation through the pathway 2 is negligible and suggests a non-slip flow. This is not surprising because graphitic surfaces are known for their lipophilicity, that is, they interact strongly with hydrocarbons. This is consistent with the recent calculation of larger interfacial friction for ethanol in graphene capillaries compared to water[32]. The non-slip behavior of organic solvents also explains why certain organic molecules (polar solvents) can uniformly intercalate between GO layers, similar to water, but their permeability remains below our detection limit. To confirm this model further, we have also ruled out the influence of polarity, dynamic diameter, and solubility parameters of organic solvents to their permeance (supplementary section 9).

Based on the understanding of organic molecule permeation through GO membranes, we propose a strategy to further improve the permeance through GO membranes without substantially reducing the organic solute rejection, even using relatively thick membranes. To this end, we used partially reduced $Mg^{2+}$ crosslinked GO membrane with 200 nm thickness, where the randomly distributed $Mg^{2+}$ ions between GO sheets play a role of spacers that introduces the disorder in the laminar structure and hence increases the permeance (supplementary Fig. 11 and supplementary section 10). These modified membranes show ~ 50% increase in permeance while keeping the dye rejection at 98% (supplementary Fig. 12). We believe that the performance of such ion-modified membranes could be further improved by optimising the cation selection for crosslinking and careful control of the reduction process.

In conclusion, we show that HLGO membranes of only several layers in thickness exhibit outstanding sieving properties accompanied by ultrafast solvent permeation. The proposed model based on non-slip permeation of organic solvents and slip-enhanced water permeation offers a long-sought explanation for the ultra-low permeability of sub-micron thick CGO membranes for organic solvents. Taking into account the excellent chemical stability of GO, the reported membrane can be used for organic solvent nanofiltration, with pharmaceutical and petrochemical industries being potential beneficiaries. The proposed strategy to enhance the nanofiltration properties of GO membranes by cation-crosslinking is enticing but further research is needed to optimize the performance.

**Methods**

**Preparation of GO membranes:** Graphite oxide was prepared by the Hummers method and then dispersed in water by sonication[1], which resulted in stable GO solutions. It is worth noting that, to avoid other possible influence, we repeatedly washed our GO nanosheets until the pH value of their solutions reaches to 7. GO membranes were prepared by vacuum filtering aqueous GO solutions through Anodisc Alumina or Nylon membrane[5] (47 mm diameter Whatman filters with 200 nm pore size). To obtain a uniform membrane, the GO suspension was diluted to less than 0.001 wt% before the vacuum filtration. After filtration, the membrane was allowed to dry under vacuum at room temperature for at least 24 hours before the measurements. The membranes with different thickness were obtained by filtrating different volume of GO suspension through the Alumina or Nylon support. It is noteworthy that the influence of the support membrane on the reported permeation were minimum due the large porosity. The solvent permeance through both the bare Alumina and Nylon support layers were found to be minimum 1000 times larger than the GO membrane on support layers suggesting hydraulic resistance from the support layers were negligible and could be ignored.

Two types of GO membranes used in this study are HLGO and CGO membranes. The difference between preparation of HLGO and CGO membrane lies in the ultrasonic exfoliation and centrifugal separation process. For HLGO membranes, the graphite oxide was exfoliated by a 3-minute ultrasonic exfoliation (40 W power) and then subsequently centrifuged twice at 3000 rpm for 10 minutes to separate un-exfoliated thick GO flakes. The supernatant GO solution was further centrifuged at 12000 rpm to separate large and small GO flakes. In this step, the sediment was collected because the small size and hence lighter GO flakes remain in the supernatant and larger GO flakes sediments. This sediment was then collected and re-dispersed in water by mild shaking and then repeated the centrifugation steps at 10000 and 8000 rpm respectively. This repeated centrifugation cycles with sequentially decreasing centrifugation speed enable the separation of medium size GO flakes from the large flakes and allows obtaining uniform large GO flakes required for the preparation of HLGO membranes. For the preparation of CGO membranes, the graphite oxide in water was sonicated for 24 hours and then centrifuged three times at 8000 rpm. The supernatant was then collected and used for the membrane preparation.

It is important to note that, the HLGO and CGO membranes are prepared by identical procedures except different exfoliation time. The influence of which on the chemical composition of GO sheets was carefully examined by X-ray photoelectron spectroscopy (XPS) and found no difference in oxygen content between two membranes (supplementary section 1 and supplementary Fig. 2).

The flake size distribution of GO used for the preparation of conventional CGO and HLGO membranes were measured by analysing more than 700 flakes with the scanning electron microscopy (SEM) or optical microscopy. Due to long time ultrasonication, all the GO flakes used for the CGO membranes are found to be smaller than 1 μm in nominal size and more than 75% of these flakes are with a size between 0.1-0.4 μm. In comparison, for HLGO membranes, 75% of the flakes used were found to be larger than 10 μm (supplementary Fig. 1).

**Membrane characterizations:** SEM and AFM techniques were used to measure the size of GO flakes and thickness of the membranes. A Veeco Dimension 3100AFM in the tapping mode was used for the AFM measurements. To measure the thickness of the GO membranes, we transferred the membrane from the alumina support to a silicon substrate by floating the alumina supported GO membrane in water and subsequently fishing out the GO membrane onto a silicon substrate. GO membrane transferred silicon substrates were completely dried in vacuum before the AFM measurements.

X-ray diffraction measurements in the $2\theta$ range of 5° to 25° (with a step size of 0.02° and recording rate of 0.2 s) were performed using a Bruker D8 diffractometer with Cu K$\alpha$ radiation ($\lambda$ = 1.5406 Å). Due to the weak intensity of the X-ray peak from an 8 nm membrane we used 70 nm thick membranes for our experiments. To collect an XRD spectrum from HLGO membranes exposed to different organic solvents, the membranes were first aged in a glovebox filled with dry argon gas for more than 5 days to remove any interlayer water present in the membranes[1,5] and then immersed in various solvents for more than 3 days inside a glove box. For the XRD measurements, the samples were collected from the solvents and kept inside an airtight XRD sample holder (Bruker, A100B36/B37) filled with same organic solvent vapour to avoid any influences of the environmental humidity and evaporation of solvent from the membrane on the measurements.

**Permeation and molecular sieving measurements:** For probing the molecular sieving and solvent permeation through various GO membranes we used a vacuum filtration setup, where the membrane is clamped and sealed with a silicone rubber O-ring between the feed and permeate side. For each test, at least three membranes were used to validate the reproducibility. Permeate side was connected to a vacuum pump with a controllable pumping speed and a cold trap. The vacuum on the permeate side creates a pressure gradient ($\Delta P$) which drives the molecular permeation across the membrane. Vacuum degree on the permeate side was controllable from 0.6 to 0.01 bar (VARIO chemistry diaphragm pump, Vacuubrand) and feed pressure was $\approx$ 1 bar. For studying the influence of $\Delta P$ on the permeance, we have performed filtration experiments with different $\Delta P$ created using different pumping speed. The permeance of various solvents was obtained by measuring both the volume and weight of the solvent from the permeate side in a liquid nitrogen cold trap and the liquid leftover in the feed side. The system leakage was examined by replacing the membrane with a 100 μm polyethylene terephthalate plastic sheet, or a 200 μm Cu foil, the leakage was found to be < 0.1 L m$^{-2}$ h$^{-1}$ bar$^{-1}$. The solvent permeance through GO membrane was also measured by a dead-end pressure filtration system at room temperature and found good consistency between two methods (Supplementary section 2).

We have noticed that for water due to its high surface tension the HLGO membrane breaks once the water was in contact with the membrane. We, therefore, used a small amount of surfactant (0.6 mg/mL sodium dodecyl benzene sulfonate) to decrease the surface tension of water and thereby avoiding the membrane damage during water permeation experiments.

For probing the molecular sieving property of HLGO and CGO membranes we used aqueous solutions of NaCl, $MgCl_2$, $K_3[Fe(CN)_6]$, pyrenetetrasulfonic acid tetrasodium salt ($Na_4PTS$), MB, RB, and BB. For MB, RB, and BB the feed concentrations were 20 mg/L, and for $K_3[Fe(CN)_6]$ and $Na_4PTS$, their concentrations were 1000, 250 mg/L, respectively. For NaCl and $MgCl_2$ we used 1M concentration. All the experiments were repeated at least three times. The amount of sodium and magnesium salts permeated were measured by probing the concentration of salt in the permeate side by checking the conductivity of the permeate water. Furthermore, we cross-checked the results of our conductivity analysis by weighing the dry material left after evaporation of water in the permeate. The permeation of other salts and dyes through GO membranes was measured by checking their concentration at the permeate side by UV-vis absorption as detailed below. The salt rejection was calculated as ($1-C_P/C_F$), where $C_P$ is the salt concentration at the permeate side and $C_F$ is the salt concentration at the feed side.

For organic solvent nanofiltration experiments, CG, MB, DR, CV, BB, and RB with a concentration of 200 mg/L were dissolved in methanol. The concentration of the dye at the permeate side was measured by UV-vis absorption as detailed below and the permeance was determined by the same method for the measurement of pure solvent as detailed above.

**UV-Vis absorption:** For obtaining the concentrations of $K_3[Fe(CN)_6]$, $Na_4PTS$ and organic dye molecules in the permeate we used optical absorption spectroscopy. UV-visible-near-infrared grating spectrometer with a xenon lamp source (240-1700 nm) was used for this study. For the HLGO membranes, we could not detect any absorption features of the above salts or dye in the permeate side (supplementary Fig. 4). To cross check this further, we have also measured the concentration of the leftover feed solution after the filtration experiment. The leftover concentrated feed solutions (including the salt or dye absorbed on the membrane) were diluted to the same volume as before the filtration experiment and then the optical absorption features were compared with the pristine original feed solution. We could not find any difference in the absorption spectra, suggesting all the solutes were retained at the feed side. The detection limit in Fig. 2a and c were estimated by measuring a reference solution and gradually decreasing its concentration until the signature peaks completely disappeared. The penultimate concentration is set as the corresponding detection limit. For the case of CGO membranes and $Mg^{2+}$-crosslinked membranes (Fig. 2a, 2c, and supplementary Fig. 12), the absorbance for the most intense optical absorption peak for various known concentrations of salt and dye molecules were plotted against their concentration and obtained a linear fit. From this linear dependence, we estimated the concentration of salt and dye at the permeate side.

# Supplementary Information

# Ultrathin graphene-based membrane with precise molecular sieving and ultrafast solvent permeation

Q. Yang, Y. Su, C. Chi, C. T. Cherian, K. Huang, V. G. Kravets, F. C. Wang, J. C. Zhang, A. Pratt, A. N. Grigorenko, F. Guinea, A. K Geim, R. R. Nair

## 1. Graphene Oxide (GO) flakes with different sizes

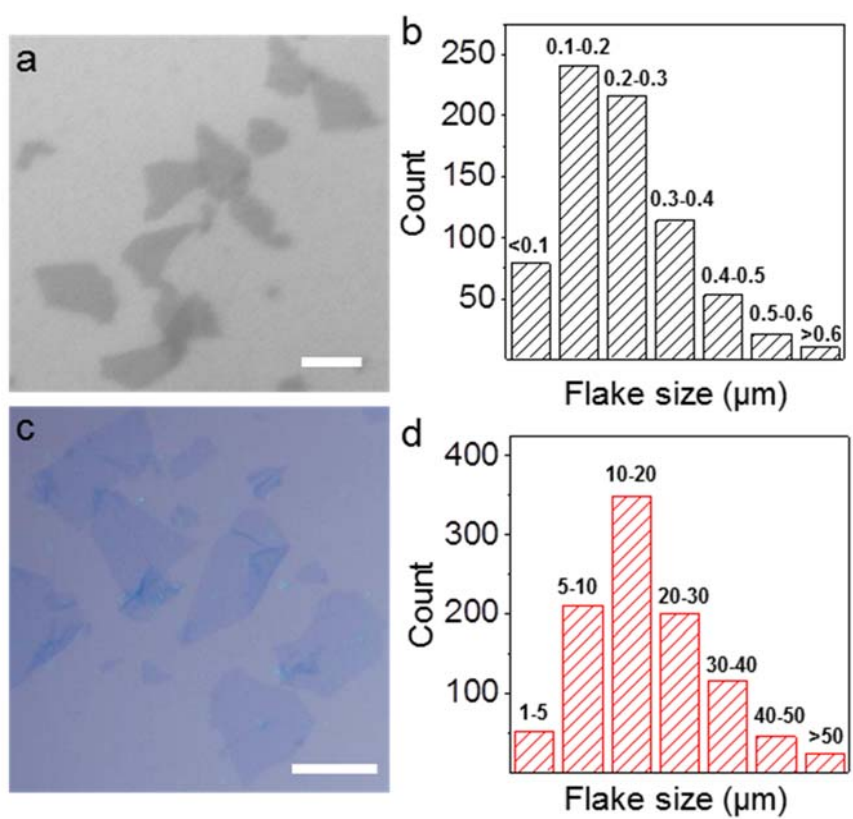

**Supplementary Fig.1| GO flake size distribution.** (a) SEM image of GO flakes used for the preparation of CGO membranes (Scale bar, 200 nm) and (b) its flake size distribution. (c) Optical image of GO flakes used for the preparation of HLGO membranes (Scale bar, 20 μm) and (d) its flake size distribution. The flake sizes were estimated by taking the square root of the area of each flake measured with the Image J software.

The influence of flake size on the chemical composition of GO sheets was carefully examined by X-ray photoelectron spectroscopy (XPS). XPS experiments were performed using a monochromated Al Kα source (1486.6 eV) in an ultrahigh vacuum system with a base pressure of $< 2\times10^{-10}$ mbar. Survey scans were taken to confirm that only C and O were present in each sample before high-resolution C 1s spectra were obtained (supplementary Fig.

2). Using XPS Peak 4.1, each C 1$s$ spectrum was fitted with four components representing the main bonding environments found in graphene oxide: C-C (284.5-284.8 eV), C-O (285.2-285.4 eV), C=O (286.8-287.2 eV), and C(=O)-(OH) (288.1-289.1 eV)[1,2]. Fitted peak areas were used to calculate C/O ratios of 3.3 ± 0.3 and 3.6 ± 0.3 for HLGO and CGO membranes, respectively. The corresponding oxygen content of 23 ± 2% for HLGO and 22 ± 2% for CGO clearly indicates that the size of GO flake does not influence their oxygen content and is consistent with the previous reports where the oxygen content is found insensitive to the GO flake size[3,4].

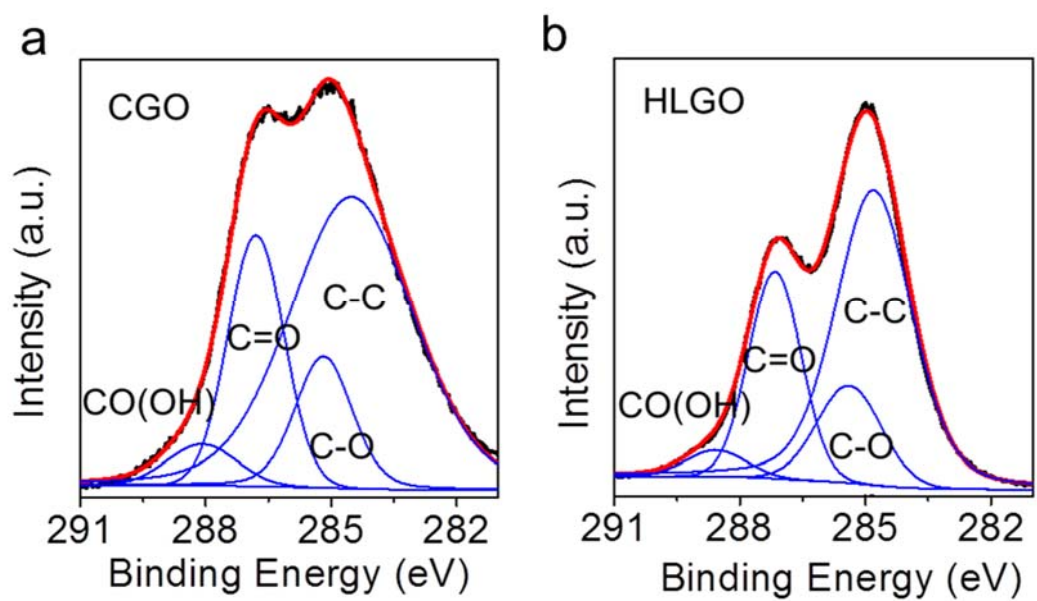

**Supplementary Fig.2| X-ray photoelectron spectroscopy of CGO and HLGO membrane.** XPS spectra from (a) CGO and (b) HLGO membrane showing raw data (black line), the fitting envelope (red line), and deconvolved peaks (blue lines) attributed to the chemical environments indicated. With respect to C 1s peak, C-C, C-O, C=O and CO(OH) peaks have an area of 61±3%, 13±2%, 22±2%, and 4±1% respectively for CGO membrane and 58±3%, 14±2%, 25±2%, and 3±1% respectively for HLGO membrane.

2. **Dead-end pressure filtration**

In addition to the vacuum filtration, HLGO membranes were also examined by pressure filtration using a home-made dead-end set-up with a pressure up to 2 bar (supplementary Fig. 3 inset). Supplementary Fig. 3a shows the methanol permeance as a function of pressure for 8, 15 and 50 nm thick HLGO membranes. As expected, the permeance was found to increase linearly with the applied pressure. 8 nm membranes were also tested for dye rejection (methylene blue, similar conditions to that of vacuum filtration) and obtained >99.9% rejection. Supplementary Fig. 3b shows the water and methanol permeance as a function membrane thickness. Similar to the case of vacuum filtration, the organic solvents' permeance decreased exponentially with the thickness whereas for water it deviates this behaviour above 50 nm. For the thicker membranes (> 50 nm) we have also used the Sterlitech HP4750 stirred cell for the dead-end filtration experiments (thinner membranes were found easy to get damaged in this cell during the sample mounting) and obtained similar results to that from the home-made pressure cell and the vacuum filtration.

We have also studied the influence of membrane thickness on the water permeance through CGO membranes and found that, even though the permeance is larger than HLGO membranes, their thickness dependent permeance follows the same trend as in the HLGO membranes (Supplementary Fig. 3b inset). Note that, despite its high permeance, the dye molecule rejection characteristics of CGO membrane were much inferior to that of HLGO membranes (Fig. 2a and c inset in main text).

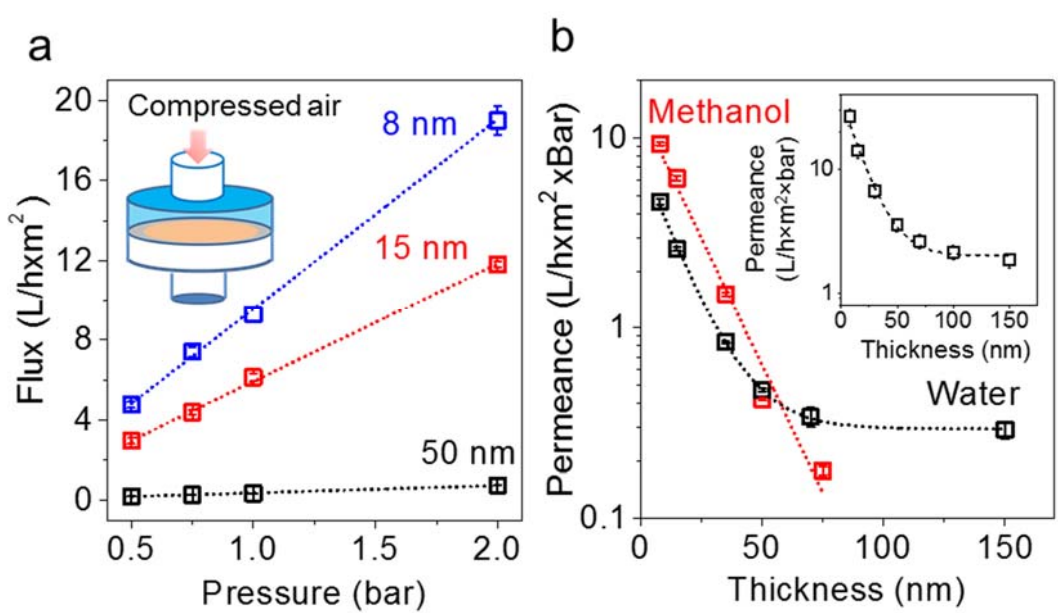

**Supplementary Fig. 3| Permeation measured by dead-end pressure filtration.** (a) Pressure dependence of methanol flux for HLGO membrane with a thickness of 8, 15 and 50 nm. Dotted lines are best linear fits. Inset: Schematic of dead-end pressure filtration setup. To avoid solvent leakage, a flat rubber gasket (marked as grey) is placed on top of the GO membrane (marked as brown), which is then clamped between two glass funnel-shaped containers. The upper compartment (marked as blue) is filled with feed solvent/solution and then controllably pressurised with compressed air. The permeate solvent/solution is collected at the bottom compartment (marked as white) and analysed as detailed in the methods session in the main text. (b) Thickness dependence of permeance for methanol and water through HLGO membranes. Inset: Thickness dependence of water permeance through CGO membranes.

## 3. Optical absorption measurements

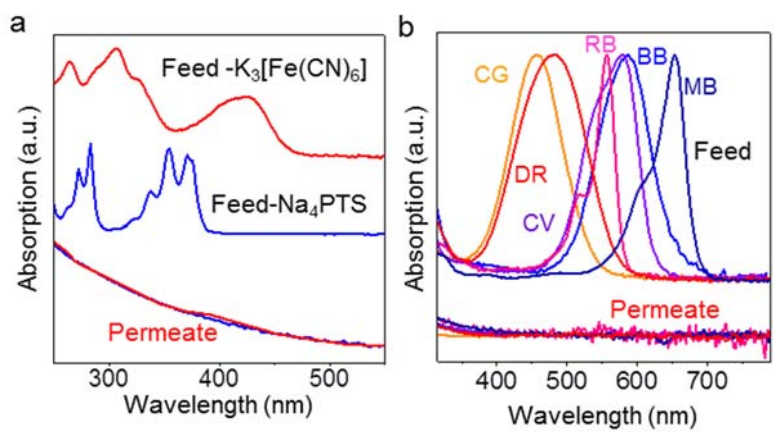

**Supplementary Fig.4| Optical detection of permeate concentration.** (a) Absorption spectra of the feed and permeate solution of $K_3[Fe(CN)_6]$ and $Na_4PTS$ in water (colour coded). (b) Absorption spectra of the feed and permeate solution of chrysoidine G (CG), disperse red (DR), methylene blue (MB), crystal violet (CV), brilliant blue (BB), and rose bengal (RB) in methanol (colour coded). The absorption spectrum from an empty container was taken as a reference spectrum of all the measurements.

## 4. Organic solvent nanofiltration (OSN) performance comparison

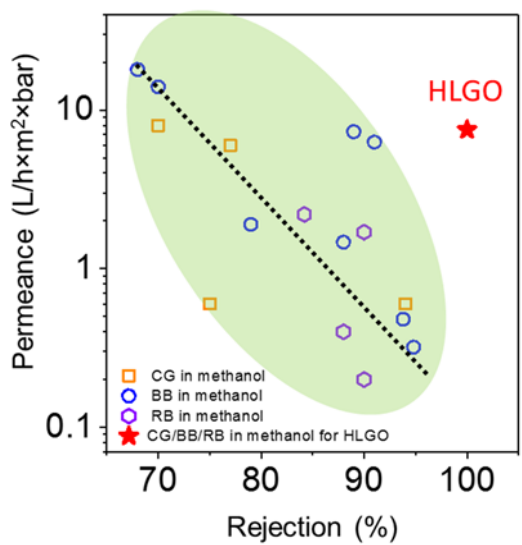

**Supplementary Fig.5| Comparison of HLGO membrane performance with other OSN membranes.** Permeance as a function of rejection for several dye molecules taken from the literature is plotted together with the data obtained from the HLGO membrane. HLGO membranes provided ≈ 100% rejection for all the tested dye molecules. Dotted line indicates the typical trend found between the rejection and permeance of reported values. All the data points in the green coloured regions are obtained from Ref. [5-10]. CG- Chrysoidine G, BB- Brilliant Blue, RB- Rose Bengal.

To further demonstrate the superior OSN performance of HLGO membranes, we have compared them with different polymeric membranes. As an example, frequently used dye molecules such as Chrysoidine G (CG), Brilliant Blue (BB), and Rose Bengal (RB) in methanol have been chosen for the comparison.

Supplementary Fig. 5 shows a typical trend between the methanol permeance and the molecular rejection values reported for several OSN membranes. Compared with the state-of-the-art polymeric membranes, HLGO membranes shows much higher permeance to solvents with a rejection of > 99.9% to dye molecules including CG whose molecular weight is only 249 g/mol. As an example, the highest reported rejection for BB in methanol is 95%, whereas the HLGO membranes exhibit a permeance of 7.5 $Lm^{-2}h^{-1}bar^{-1}$ (25 times higher) with ≈ 100% rejection. The high permeance along with ≈100% rejection even for smaller molecules indicates the prospect of HLGO membranes for OSN technology.

## 5. HLGO membrane on porous nylon support

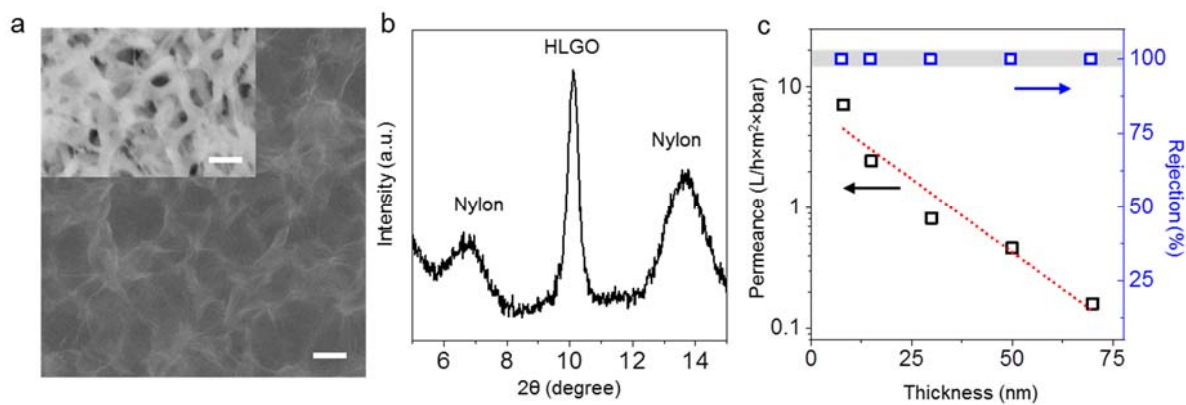

**Supplementary Fig.6| Ultrathin HLGO membrane on nylon support.** (a) SEM image of an 8 nm HLGO membrane on a nylon support. Scale bar, 1 μm. Inset: SEM image of a bare nylon support. Scale bar, 1 μm. (b) Ambient air XRD spectrum for HLGO membrane on nylon support. The peaks at ~7° and 14° are from the nylon support. (c) Permeance and rejection of MB in methanol through HLGO membranes with different thicknesses on nylon support. The dotted line is the best linear fit. Points within the grey bar show the rejection estimated from the detection limit.

In addition to the porous alumina support, which is brittle, we have also tested porous polymer as a support material. It has been reported that due to the roughness and non-uniform macroscopic pore distribution of polymer support, tens of nanometre thin GO membrane (small GO flakes) fails to maintain a good laminar structure[11]. Here, we show that GO membrane prepared from large GO flakes could form a good laminate, even though the membrane is ultrathin. Supplementary Fig. 6a shows the SEM image of a bare nylon support and an 8 nm HLGO membrane deposited nylon support. X-ray diffraction (XRD) spectrum of a 50 nm HLGO membrane on nylon substrate shows a narrow peak with a full width at half maximum (FWHM) of 0.4 degree (supplementary Fig. 6b), which confirms the highly laminated structure similar to that on the alumina support. To evaluate the organic solvent nanofiltration (OSN), we have tested filtration of methanol solutions of CG and MB through an 8 nm thin HLGO membrane on nylon support. Similar to that of alumina support, HLGO membrane on nylon support also shows a 99.9% rejection to CG and MB with a similar methanol permeance to that of alumina support (supplementary Fig. 6c). Also, the exponential decay of the methanol permeance (supplementary Fig. 6c) with increasing the thickness of HLGO membrane is consistent with that of the alumina supported HLGO membranes (Fig 3b in main text).

### 6. **Stability of HLGO membranes**

To study the stability of HLGO membrane in air and solvents, we have conducted two sets of experiments. First, for probing air stability, we have compared methanol and water permeance of freshly prepared membranes with aged membranes (two samples aged for 452 days). Both the membranes provided similar permeance indicating no significant degradation of the membrane with aging. This is consistent with ref. [12], where only ~ 6% reduction in oxygen content is reported with GO aging for 100 days. This small change in oxygen content is not expected to affect the membrane performance because the molecular permeation mainly occurs through the pristine graphene capillaries in the GO membrane and moreover, in HLGO membranes, permeation of solvents mainly occurs through the random pin holes and that is not anticipated to change with aging.

Secondly, for probing the membrane stability in the solvent environment, we have performed long time filtration experiments. Nylon supported HLGO membranes with thicknesses of 8 nm, and 30 nm were examined under dead end filtration for water and organic solvents. As shown in supplementary Fig. 7, permeance of the solvents and water are stable within the testing period which varies from 5 hours to 4 days, suggesting that the HLGO membrane is intact and capable of long-time filtration process. To further check the solvent stability of membranes we have immersed 50 nm thin HLGO membranes on nylon in water, methanol, and hexane for 7 days. Note that the 50 nm thick samples are chosen because the membrane thinner than that gives poor visual contrast for analysis. As shown in supplementary Fig. 7, all the membranes were found stable in all tested solvents and water. To quantify this further, we have measured methanol permeance before and after immersing the membrane in solvents and could not find any detectable change. To demonstrate the solvent stability of the membrane further, we have performed another vigorous testing. Membranes immersed in solvents for 7 days were further placed in a glass beaker containing the same solvent and bubbled with nitrogen gas to check if the harsh dynamic turbulence would destroy the membrane. Membranes were found intact even in such a harsh environment. This is consistent with our previous report were membranes also found stable under mild sonication[13]. We do notice that there is a debate on the stability of GO membrane in water and the currently proposed mechanism of stability is cross-linking of GO flakes with metal ions from the support substrate[14]. However, we believe many other factors could also lead to stability. For example, we found that complete drying of the membrane after its fabrication is critical for obtaining a stable membrane.

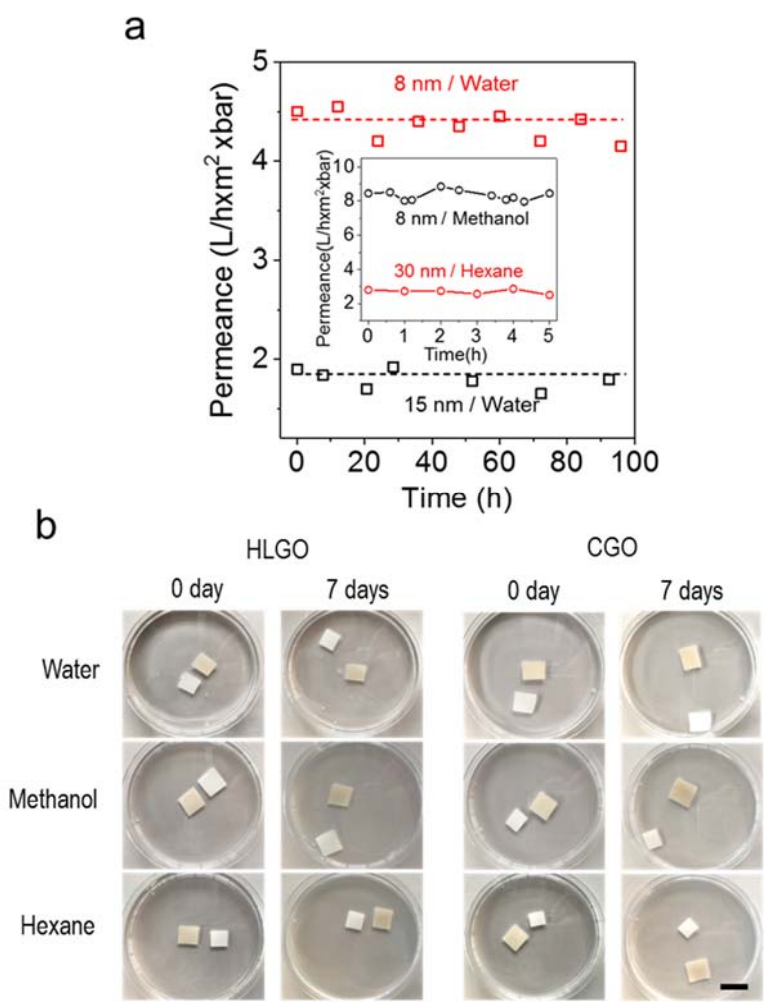

**Supplementary Fig. 7| Stability of GO membranes.** (a) Variation of water and solvent (inset) permeance through HLGO membranes as a function of measurement time. (b) Photographs showing 50 nm thin HLGO and CGO membranes on nylon support immersed in water and solvents for 0 day and 7 days. In each experiment, a bare nylon substrate (white colored substrate in the photo) was also immersed as a reference to check visual contrast of thin GO membranes on nylon. Scale bar: 15 mm.

7. <u>**Vapour and Helium gas permeation through HLGO membranes**</u>

Besides liquid permeation, vapour and gas (helium) permeation through HLGO membranes with different thicknesses ($h$) were measured to further validate the proposed mechanism for molecular transport in GO membranes. The vapour permeation measurements were performed as we reported previously[15]. Membranes were glued to a Cu foil with an opening of 0.5 cm in diameter. The foil was then clamped between two rubber O-rings sealing a metal container. Permeation was measured by monitoring the weight loss (for ≈ 12 hours) of the container that was filled with water and isopropyl alcohol (IPA) inside a glovebox. Supplementary Fig. 8a shows the weight loss rate for water and IPA through HLGO membranes with different thicknesses. Weight loss rate for IPA was found to decay exponentially with increasing membrane thickness, indicating exponentially decaying permeance, consistent with the mechanism proposed (permeation through pinholes) in the main text. However, for water, we observed a thickness independent weight loss rate, consistent with the previous report[15]. In this case, unlike liquid permeation reported in the main text, water vapour permeation is limited by the evaporation from the top surface of GO membranes and hence masks the thickness dependence.

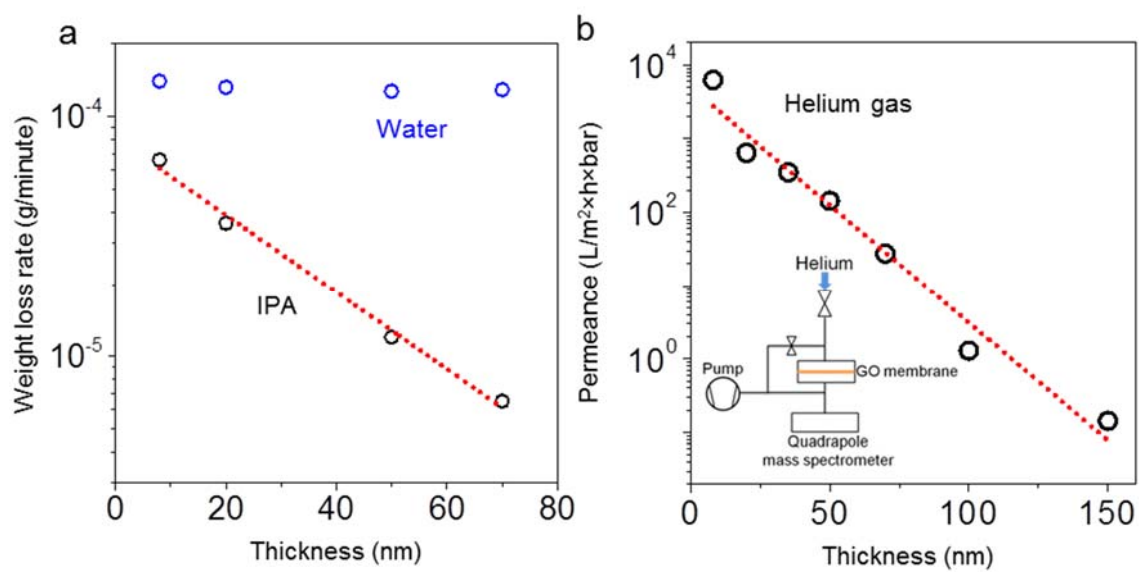

**Supplementary Fig. 8| Vapour and gas permeation through HLGO membranes.** (a) Weight loss rate for a container sealed with HLGO membranes with different thicknesses (aperture diameter ≈ 0.5 cm). Weight loss for IPA and water were tested at room-temperature and zero humidity. (b) Thickness dependence of helium permeance through HLGO membrane. Dotted lines are the best fits to the exponential decrease. Inset: Schematics of our experimental setup for helium permeation measurement.

For the helium (He) gas permeation experiments, HLGO membranes attached to the Cu foil were placed between two rubber O-rings in a custom made permeation cell and pressurised from one side up to 100 mBar. He gas permeation through the HLGO membrane was monitored on the opposite (vacuum) side by using mass spectrometry (supplementary Fig. 8b inset). We used Hiden quadrupole residual gas analyser for measuring the partial pressure of He gas in the vacuum side. A standard calibrated leak (Open style CalMaster Leak Standard, LACO technologies) is utilised to convert the partial pressure to the leak rate[16].

Supplementary Fig. 8b shows the He permeance through HLGO membrane as a function of membrane thickness. Similar to the organic solvent and vapour permeation (Fig. 3b and supplementary Fig. 8a), He gas also follows exponential decay indicating the pathway for the gas permeation is dominated by the pinholes. The observed exponential decay of He permeance with increasing thickness is consistent with the earlier study on He and $H_2$ permeance through ultrathin GO membranes[17], but the mechanism of exponential dependence was not elucidated. The proposed mechanism in this study (main text) clarifies this ambiguity.

## 8. Pinholes in ultrathin HLGO membranes

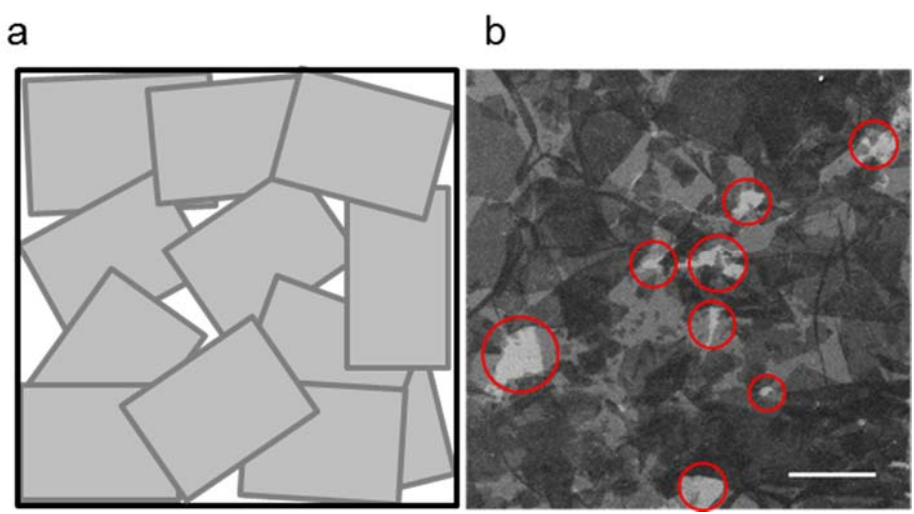

**Supplementary Fig. 9| Pinholes in GO membrane.** (a) Schematic showing continuous interconnected GO plane formed by the random overlap of GO flakes. (b) SEM image from one of our HLGO membrane with a thickness of ≈ 3 nm transferred to ITO coated glass slide showing the presence of pinholes (large pinholes are circled) in the membrane. Scale bar, 20 μm. The membrane was transferred to ITO substrate by floating the alumina supported GO membrane in water and subsequently fishing out the GO membrane onto an ITO substrate. ITO substrate was used to avoid the charging effect during SEM imaging.

During the self-assembly of GO membrane, the flakes randomly overlap and provide a continuous interconnected plane that contains a large number of holes (Supplementary Fig. 9). These holes between different flakes are referred as pinholes. Our SEM analysis shows that for ≈ 3 nm membranes the size of these pinholes is of the order of the flake size. With increasing numbers of layers of GO, the newly added layers block these pinholes and form fully covered GO membranes. Our sieving experiments (Fig. 1) confirm that the minimum thickness required for the fully continuous GO membrane is ~ 8 nm.

## 9. Influence of solvent parameters on their permeance

To probe other possible mechanisms (e.g. Solution-diffusion model[18]) for the faster water transport through thick GO membranes, we have checked the correlation between different solvent parameters such as relative polarity, kinetic diameter and total Hansen solubility parameter[19-26] on the permeance through the HLGO membranes. Supplementary Table 1 shows the different parameters for the solvents used in our experiments. To understand the influence of the above parameters on the permeance we have plotted the product of permeance and viscosity as a function of the solvent parameters. As an example, supplementary Fig. 10 shows Hansen solubility parameter vs. product of viscosity and permeance for 8 nm and 70 nm thick HLGO membranes. Despite a small variation (within the grey area in supplementary Fig. 10), it is clear from the figure that the permeance behaviour for 8 nm HLGO membrane is very close to what can be expected for pore flow model[27], i.e. product of viscosity and permeance is a constant, independent of the solvents used. On the other hand for 70 nm thick membranes, except water, for all other solvents, the product of permeance and viscosity is nearly a constant. However, water deviates from this trend and permeates faster. A similar behaviour can also be obtained by plotting product of viscosity and permeance as a function other solvent parameters. The unique fast permeation of water through GO membranes is attributed to the enhanced flow of water through graphene capillaries in the GO membranes[15]. By increasing the GO membrane thickness further, we only obtained water flux and all the organic solvent flux were below our detection limit. The absences of correlation between solvent parameters and permeance of HLGO membranes, especially for the thicker membranes (> 50 nm), further support the validity of pore flow model for mass transport in GO membranes and rules out other mechanisms such as solution-diffusion model[18].

**Supplementary Table 1| Solvent parameters.** Viscosity, relative polarity, kinetic diameter, and total Hansen solubility parameter of the solvents used in this study.

| Solvents | Viscosity[19,20] | Relative polarity[21,22] | Kinetic diameter[23] | Total Hansen solubility parameter[24,25,26] |
|---|---|---|---|---|
| | mPa.s | | nm | MPa$^{1/2}$ |
| hexane | 0.3 | 0.009 | 0.51 | 14.9 |
| acetone | 0.306 | 0.355 | 0.47 | 20.1 |
| actonitrile | 0.369 | 0.46 | 0.34 | 24.4 |
| tetrahydrofunan | 0.456 | 0.207 | 0.48 | 19.4 |
| methanol | 0.544 | 0.762 | 0.38 | 29.7 |
| butyl acetate | 0.685 | - | - | 16.8 |
| water | 0.89 | 1 | 0.265 | 47.8 |
| ethanol | 1.074 | 0.654 | 0.44 | 26.6 |
| iso-proponal | 2.038 | 0.546 | 0.47 | 24.6 |
| butanol | 2.544 | 0.586 | 0.5 | 23.1 |

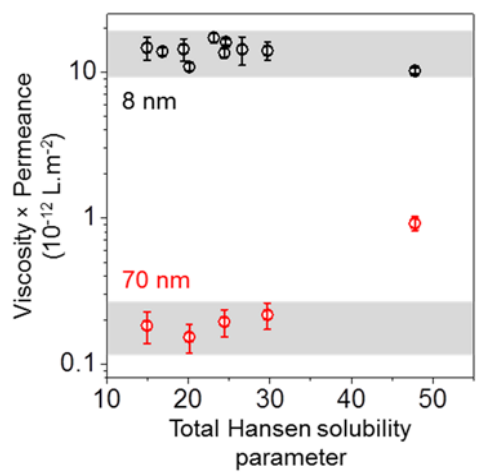

**Supplementary Fig. 10| Probing influence of solubility parameter of solvents on permeation.** Product of permeance and viscosity of solvents as a function of the total Hansen solubility parameter for 8 nm and 70 nm thick HLGO membrane (colour coded). Variations between the points in the grey colour marked region are minor and within the accuracy of measurements. All the error bars are standard deviations using at least three different measurements using different samples.

## 10. $Mg^{2+}$-crosslinked partially reduced GO membrane for OSN

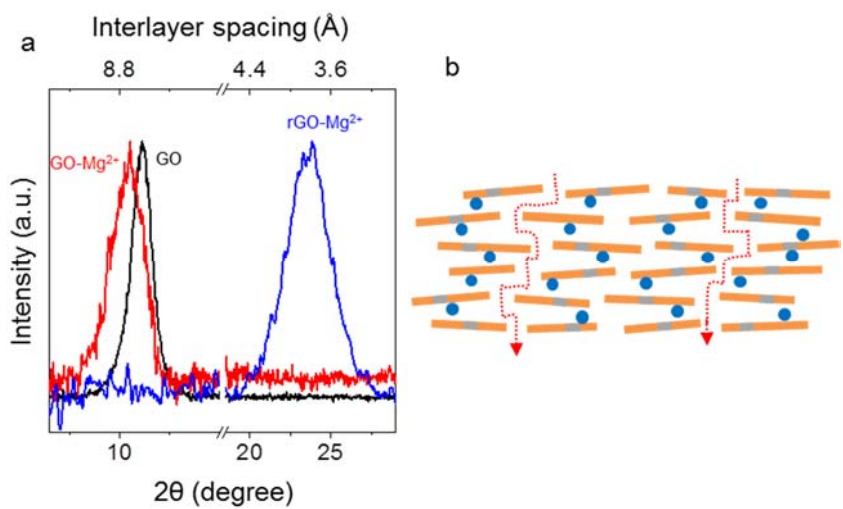

**Supplementary Fig. 11| $Mg^{2+}$ crosslinked GO membranes.** (a) X-ray diffraction for pristine GO, $Mg^{2+}$ crosslinked GO (GO-$Mg^{2+}$) and partially reduced $Mg^{2+}$ crosslinked GO (rGO-$Mg^{2+}$) membranes. The thickness of membranes ≈ 200 nm. (b) Schematic showing the structure of the GO-$Mg^{2+}$ membrane. The dotted line indicates the permeation pathway and blue circles indicate $Mg^{2+}$ ions.

Multivalent cations have previously been used to crosslink the GO sheets by attaching them to the oxidised regions to improve the mechanical strength and to control the ion permeation through the GO membranes[14,28,29]. Here, we propose the same crosslinking technique to enhance the solvent permeance through the GO membranes because the interlayer cations could act as randomly distributed external spacers to introduce disorder in the laminar structure (supplementary Fig. 11) and hence increase the permeance. We chose $Mg^{2+}$ for crosslinking due to its large hydrated diameter[13], which is comparable to the interlayer spacing in GO membrane.

GO crosslinking with $Mg^{2+}$ was carried out by the drop-by-drop addition of 10 mL of 9.5 g/L $MgCl_2$ into 40 mL GO suspension (0.2 wt. %) under vigorous magnetic stirring followed by at least one day of sonication. After the sonication, the suspensions were stable up to one hour (average flake size ≈ 200 nm) without any stirring, but it starts agglomerating after that. This could be due to the neutralisation of the negative surface charges of GO with the cations. To avoid the agglomeration we stored the suspension under vigorous stirring. $Mg^{2+}$ crosslinked GO membranes (GO-$Mg^{2+}$) were then prepared by the vacuum filtration of these suspensions through an Anodisc alumina membrane (200 nm pore size). The incorporation of $Mg^{2+}$ in the GO membranes was confirmed by XRD analysis, where a broader GO peak was found (supplementary Fig. 11a). An increase of FWHM from 1.6 degree to 2.1 degree indicates a poor interlayer alignment in GO-$Mg^{2+}$ (supplementary Fig. 11b) compared to pristine GO and suggests the prospect of obtaining higher permeance. The organic solvents permeance and organic solvent nanofiltration (OSN) through GO-$Mg^{2+}$ membranes (200 nm thick) were measured by vacuum filtration technique as detailed in the main text. Supplementary Fig. 12 shows the pure solvent permeance and dye rejection properties of GO-$Mg^{2+}$ membranes. Comparing to the performance of the CGO membranes, even though GO-$Mg^{2+}$ membranes are thicker, they show nearly one order of magnitude higher permeance to methanol but with same dye rejection (84% MB rejection for 35 nm CGO and 200 nm GO-$Mg^{2+}$ membrane) (supplementary Fig. 12b and Fig. 2c inset). The enhanced permeance through GO-$Mg^{2+}$ membranes suggests that the addition of $Mg^{2+}$ increases the disorder in the laminar structure as shown in supplementary Fig. 11b.

To further improve the dye rejection performance of the GO-$Mg^{2+}$ membranes, we partially reduced them in hydroiodic acid vapour for 1 min at room temperature. The partially reduced GO-$Mg^{2+}$ membranes (rGO-$Mg^{2+}$ membranes) show a broad XRD peak at ≈ 23.7º (supplementary Fig. 11a), suggesting the collapse of the interlayer channels. However, in comparison to the fully reduced GO membranes[30] (peak at ≈ 25º with FWHM of 1.7 degree), where it blocks the permeation of all gases and solvents, the larger FWHM of 3.3 degree confirms larger disorder in the laminar structure which could allow the molecular permeation. Our filtration experiments further support this. After the partial reduction, even though the permeance of all the solvents decreased by a factor of ≈ 3.5 (supplementary Fig. 12a), it is still 30-50% higher than even the permeance for an 8 nm thick HLGO membranes. Besides, the rGO-$Mg^{2+}$ membranes exhibited 90-99% rejections to the organic dye molecules with molecular weights ranging from 249 g/mol to 1017 g/mol (supplementary Fig. 12b). We explain the relatively lower permeance and high rejection of dye molecules for rGO-Mg$^{2+}$ membranes compared to GO-Mg$^{2+}$ membranes by the close packing[30] of the interlayer after the reduction, which could make the disordered interlayer channels narrower. Even though further improvement in membrane performance could be achieved with better optimisations in the membrane crosslinking process, our findings show the potential of crosslinked GO membrane for organic solvent nanofiltration applications.

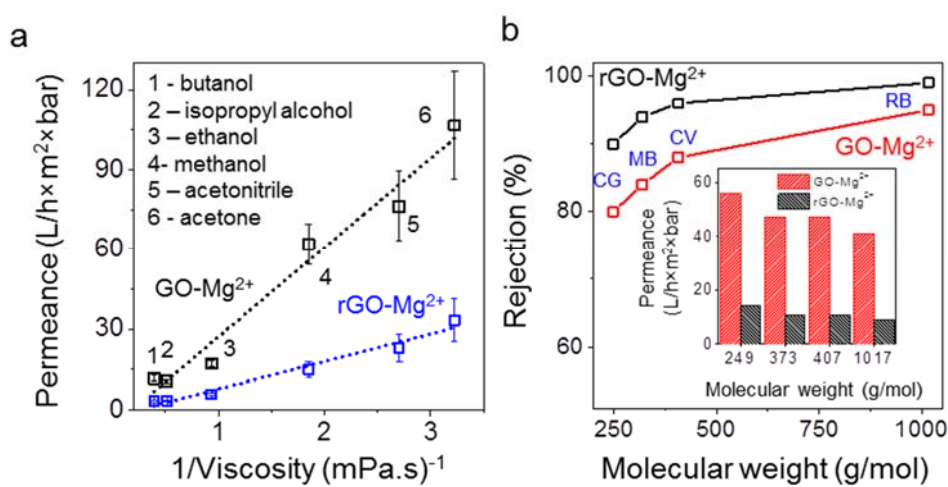

**Supplementary Fig. 12| Permeation through 200 nm thick Mg$^{2+}$-crosslinked GO membranes.** (a) Permeance of various organic solvents through GO-Mg$^{2+}$ and rGO-Mg$^{2+}$ membrane as a function of their inverse viscosity. The used solvents are numbered and named on the top left. Dotted lines are the best linear fit. (b) Rejection of several dyes in methanol versus their molecular weight. The dyes used: CG, MB, CV, and RB. Inset: The corresponding permeance of methanol.

### Supporting References